\documentstyle{article}
\begin{document}
\author{Anatoly Konechny and  Albert Schwarz\\
 \\
Department of Mathematics, University of California \\
Davis, CA 95616 USA\\
konechny@math.ucdavis.edu,  schwarz@math.ucdavis.edu}
\title{\bf Supersymmetry algebra and BPS states of
super Yang-Mills theories on   noncommutative tori }
\date{January 15, 1999}
\maketitle
\begin{abstract}
We consider 10-dimensional super Yang-Mills theory with topological terms
compactified on a noncommutative torus.
We calculate supersymmetry algebra and derive BPS energy spectra from it.
The cases of $d$-dimensional tori with $d=2,3,4$ are considered in full detail.
$SO(d,d,{\rm \bf Z})$-invariance of the BPS spectrum and relation of new
results to the
previous work in this direction are discussed.
\end{abstract}
\large

\section{Introduction}
In this paper we consider supersymmetric gauge theories  on
noncommutative tori.
It was shown in \cite{CDS} that such gauge theories can be obtained as
compactifications of M(atrix) theory
. More precisely we can consider compactifications of BFSS Matrix model
(\cite{BFSS}) on a noncommutative $d$-dimensional torus $T_{\theta}$.
Another way to look at this model
is by considering first a compactification of IKKT model \cite{IKKT} 
(as it was suggested in \cite{CDS})
on a $(1+d)$-dimensional torus of the form $T=S^{1}\times T_{\theta}$ and
then
performing a Wick rotation of that theory with the $S^{1}$ factor corresponding to the time direction.

For the case of commutative torus our calculations agree with well known results (see for example
\cite{OP}). They agree also with calculations performed in \cite{HofVer2} in the context of Born-Infeld
theory. Partial answer to the problem we consider is contained also in \cite{BrMor}, \cite{BrMorZum2}. 
The terms that were taken into account in these papers agree with our results.

The paper is organized as follows. In section 2 we calculate supersymmetry algebra of the model and
derive the expression for BPS energy spectrum out of it. In section 3 we calculate BPS spectra for
$d=2,3,4$ explicitly in terms of topological numbers and eigenvalues of the total momentum and
the zero mode of electric field. In section 4 we discuss the invariance of the BPS spectra with respect
to $SO(d,d|{\rm \bf Z})$ transformations. Finally, we discuss the case of modules admitting a constant curvature connection. We show that for this case our new results can be  reproduced by the method used in our
previous paper \cite{KS} for the analysis of modules admitting constant curvature connection over 
two-dimensional tori. 

\section{Supersymmetry algebra and BPS spectrum}

Let  $T_{\theta}$ be a  $d$-dimensional noncommutative torus specified by an antisymmetric
$d\times d$ matrix $\theta$. Such a torus can be defined by means of generators
$U_{1}, \dots , U_{d}$ satisfying
$$
U_{k}U_{j} = e^{2\pi i\theta^{kj}} U_{j}U_{k} \, .
$$
Let us consider derivations $\delta_{i}$ of $T_{\theta}$  determined by the following relations
\begin{equation} \label{delta}
\delta_{k} U_{j} = 2\pi i\delta_{kj} U_{j}
\end{equation}
where $\delta_{kj}$ is the Kronecker symbol. These derivations span an abelian Lie algebra denoted
 by $L_{\theta}$.

Consider  an abelian 10-dimensional Lie algebra $L$
that acts on $T_{\theta}$ by means of derivations belonging to $L_{\theta}$.
This means that for every $X\in L$ we have an operator
$\delta_{X}\in L_{\theta}$ obeying $\delta_{X}(ab) = \delta_{X}(a) b +
a\delta_{X}(b)$ for any $a,b \in T$. 
Then a connection $\nabla_{X}$ on $E$ can be defined as
a set of linear operators $\nabla_{X}: E\to E$, $X\in L$ satisfying the Leibnitz rule:
\begin{equation} \label{Leib}
\nabla_{X}(ae) = a\nabla_{X}e + (\delta_{X}a)e
\end{equation}
for any $a\in T$, $e\in E$. (We assume that $\delta_{X}$ and $\nabla_{X}$ depend linearly 
on $X$.)
We also suppose that  $L$ is equipped with a Minkowski signature metric.
These data are sufficient to define an action functional of the
M(atrix) theory (BFSS model) compactified on $T_{\theta}$.

Let us adopt the following conventions about indices. The capital Latin
indices $I,J, \dots$  run from 0 to 9, Greek indices $\mu$, $\nu, \dots$
run from 0 to $d$,  the small Latin indices $i,j, \dots$  take values
from 1 to $d$, and the tilted small Latin indices $\tilde\imath, \tilde\jmath, \dots$
take values from $d+1$ to $9$.
We assume that one can fix a basis  $X_{I}$ of the Lie algebra $L$ such that 
$\delta_{X_{i}}=\delta_{i}$, $\delta_{\tilde \imath}=0$, and the metric tensor 
 $g_{IJ}$ in this basis satisfies   $g_{00} = -R_{0}^{2}$, $g_{I0} = 0$  if
$I\ne 0$, $g_{i\tilde\jmath}=0$, $g_{\tilde\imath, \tilde\jmath} =
\delta_{\tilde\imath, \tilde\jmath}$.
The conditions on $g_{IJ}$ mean that it can be written as the following block
matrix
\begin{equation}
g = \left(
\begin{array}{ccccc}
-R_{0}^{2} & 0 & 0 & \dots & 0 \\
0 & (g_{ij} ) & \dots &  0 & 0 \\
0& \dots & 1 & \dots & 0 \\
0& \dots & \dots& \ddots & 0 \\
0& 0 & \dots & \dots & 1\\
\end{array} \right)
\end{equation}
where $g_{ij}$ stands for a $d\times d$ matrix that defines a metric
on the spatial torus.

The Minkowski action functional of M(atrix) theory compactified on a noncommutative
torus $T_{\theta}$ can be written as
\begin{eqnarray} \label{action}
S =&& \frac{-V}{4g_{YM}^{2}} {\rm Tr} (F_{IJ} + \phi_{IJ}{\bf 1})
( F^{IJ} + \phi^{IJ}{\bf 1}) + \nonumber \\
&& + \frac{iV}{2g_{YM}^{2}} {\rm Tr} \bar \psi \Gamma^{I} [\nabla_{I} ,
\psi ]
\end{eqnarray}
where  $\nabla_{I}$ is a connection on a $T_{\theta}$-module $E$, 
$\psi$ is a ten-dimensional Majorana-Weyl spinor taking values in the algebra $End_{T_{\theta}}E$ 
of endomorphisms of $E$, $\phi_{IJ}$ is an antisymmetric tensor with  the only
non-vanishing components $\phi_{ij}$, and $\Gamma_{I}$ are gamma-matrices satisfying
$\{ \Gamma_{I}, \Gamma_{J}\} = -2g_{IJ}$. It follows from our conventions on $\delta_{X_{I}}$
that $\nabla_{\tilde \imath} = X_{\tilde \imath}$ where $X_{\tilde \imath}$ are endomorphisms 
of $E$ (scalar fields).


In \cite{KS} we discussed the quantization  of  Yang-Mills theory on a
noncommutative torus in the $\nabla_{0} = 0$ gauge.
Those considerations can be easily generalized to the supersymmetric case
at hand. Let us briefly describe the quantization procedure. The Minkowski
action functional (\ref{action}) is defined on the configuration space 
$ConnE\times (\Pi End_{T_{\theta}}E)^{16}$
where $ConnE$ denotes the space of connections on $E$, $\Pi$ denotes the
parity reversion operator. To describe the Hamiltonian formulation we
first restrict ourselves
to the space
${\cal M} = Conn' E\times (End_{T_{\theta}}E)^{9} \times (\Pi
End_{T_{\theta}}E)^{16}$
where $Conn'E$ stands for the space of connections satisfying
$\nabla_{0}=0$, the second factor
corresponds to a cotangent space to $Conn'E$. We denote coordinates on that cotangent 
space by $P^{I}$. 
Let ${\cal N}\subset {\cal M}$ be a subspace where the Gaussian constraint $[\nabla_{i},P^{i}]=0$
is satisfied.  
Then the  phase space of the theory is the quotient ${\cal P}={\cal N}/G$ where $G$
is the  group of spatial gauge transformations.

The presymplectic form (i.e. a degenerate closed 2-form)  $\omega$ on ${\cal M}$ is defined as
\begin{equation} \label{sform}
\omega = {\rm Tr} \delta P^{I}\wedge \delta \nabla_{I}
-i\frac{V}{2g_{YM}^{2}}{\rm Tr}\delta \bar \psi \Gamma^{0} \wedge \delta \psi \, .
\end{equation}
It descends to a symplectic form on the phase space $\cal P$ which determines Poisson brackets
$\{. , . \}_{PB}$.

The Hamiltonian corresponding to (\ref{action}) reads
\begin{eqnarray}\label{Hamiltonian}
&&H= {\rm Tr} \frac{g_{YM}^{2}R_{0}^{2}}{2V}
P^{I}g_{IJ}P^{J}  + \nonumber \\
&& +  {\rm Tr}\frac{V}{4g_{YM}^{2}}(F_{IJ} +
\phi_{IJ}\cdot {\bf 1})g^{IK} g^{JL}(F_{KL} +
\phi_{KL}\cdot {\bf 1}) +  \nonumber \\
&& + \mbox{fermionic term} \, .
\end{eqnarray}

The  action (\ref{action}) is invariant
under the following supersymmetry
transformations
\begin{eqnarray} \label{susy}
&&\delta_{\epsilon} \nabla_{I} = \frac{i}{2}\bar \epsilon \Gamma_{I}\psi
\nonumber \\
&& \delta_{\epsilon} \psi = -\frac{1}{4} F_{IJ}\Gamma^{IJ}
\epsilon \\
&& \tilde \delta_{\epsilon} \psi = \epsilon \, , \quad
\tilde \delta_{\epsilon} \nabla_{I} = 0
\end{eqnarray}

The corresponding supercharges  are given by expressions
\begin{equation} \label{sch1}
 Q = \frac{iR_{0}}{2}{\rm Tr} P^{I}\psi^{t} \Gamma_{0} \Gamma_{I} +
\frac{iVR_{0}}{4g_{YM}^{2}}{\rm Tr}F_{IJ}\psi^{t}  \Gamma^{IJ} \, ,
\end{equation}
\begin{equation}\label{sch2}
 \tilde Q = -\frac{iVR_{0}}{g_{YM}^{2}}{\rm Tr}\psi \, .
\end{equation}

( Supersymmetry transformations are odd vector fields preserving the
symplectic form and therefore are generated by odd functions on the phase
space - supercharges.)

As one can readily calculate using (\ref{susy}) and (\ref{sch1}), (\ref{sch2}) the supersymmetry
algebra has the form
\begin{eqnarray} \label{salgebra}
&&i \{Q_{\alpha}, Q_{\beta} \}_{PB} = \frac{1}{2} (H - \Phi )
+  \frac{R_{0}}{2}\gamma^{I}{\cal P}_{I} -\frac{V}{16g_{YM}^{2}}
\gamma^{[ijkl]} C_{ijkl}  \nonumber \\
&&i \{\tilde Q_{\alpha}, Q_{\beta} \}_{PB} = \frac{-R_{0}}{2} \gamma_{I} p^{I}
+ \frac{V}{4g_{YM}^{2}}\gamma^{ij}C_{ij} \nonumber \\
&&i \{ \tilde Q_{\alpha}, \tilde Q_{\beta} \}_{PB} = \frac{V}{g_{YM}^{2}}
dimE
\end{eqnarray}
where $H$ is the Hamiltonian (\ref{Hamiltonian}),
$\Phi = \frac{V}{4g_{YM}^{2}}{\rm Tr}(\phi_{ij}\phi^{ij}{\bf 1} + 2F_{ij}\phi^{ij})$,
$p^{I} = {\rm Tr}P^{I}$,
 ${\cal P}_{I} = {\rm Tr}F_{IJ}P^{J} + \mbox{fermionic term}$,
$C_{ij} = {\rm Tr}F_{ij}$, $C_{ijkl}= {\rm Tr} F_{ij} F_{kl}$.
In (\ref{salgebra}) we  use the Euclidean nine-dimensional gamma-matrices
$\gamma^{I} = R_{0} \Gamma^{0}\Gamma^{I}$ satisfying
$\{ \gamma^{I}, \gamma^{J} \} = 2g^{IJ} $. Note that in (\ref{salgebra}) we assume
that the index $I$ does not take the zero value. When calculating Poisson brackets (\ref{salgebra})
it is convenient to use the identity $\{ A, B \}_{PB} = \frac{1}{2}({\bf a}(B) - {\bf b}(A))$ where
$\bf a$, $\bf b$ denote the Hamiltonian vector fields corresponding to functions $A$ and $B$
respectively.

After quantization the supersymmetry algebra (\ref{salgebra}) preserves its form (one only needs
to drop factors of $i$ on the LHS's), provided $Q_{\alpha}$, $\tilde Q_{\alpha}$, $H$, ${\cal P}_{I}$,
 and $p^{I}$  are considered as self-adjoint operators in a Hilbert space. The quantities $C_{ij}$,
$C_{ijkl}$, $\Phi$, $dimE$ are central charges.
As usual we define BPS states as states annihilated by a part of supersymmetry operators.
The energy $H$ eigenvalues of BPS states can be expressed via
values of central charges and eigenvalues of operators ${\cal P}_{I}$, $p^{I}$ (the operators
$H$, ${\cal P}_{I}$, and $p^{I}$ all commute with each other).
The energy of  a 1/4-BPS state on a $d$-dimensional torus for $d\le 4$ is given by the formula
\begin{eqnarray} \label{energy}
&&E=\frac{R_{0}^{2}g_{YM}^{2}}{2VdimE}p^{I}g_{IJ}p^{J} + \nonumber \\
&& + \frac{V}{4g_{YM}^{2}dimE}(C_{ij}+ dimE\phi_{ij})(C^{ij} + dimE\phi^{ij}) + \nonumber \\
&&+ R_{0}\sqrt{\|{\bf v}\|^{2} + {\cal P}_{\tilde\imath}^{2}
+ (\pi /g_{YM})^{4}(C_{2})^{2} } \, .
\end{eqnarray}
where $\|{\bf v}\|^{2}= v_{i}g^{ij}v_{j}$ stands for the norm squared of  a $d$-dimensional vector
  ${\bf v} = ({\cal P}_{i} - (dimE)^{-1}C_{ij}p^{j})$, and
$$
C_{2}=\frac{1}{8\pi^{2}} \epsilon^{ijkl}(C_{ijkl} - (dimE)^{-1}C_{ij}C_{kl}) \, .
$$ In formula (\ref{energy}) we
use the same notations ${\cal P}_{I}$, $p^{I}$ for the eigenvalues of the corresponding operators.
Note that  when the term with the square root
vanishes we get a 1/2-BPS state.
Here we would like to make few remarks on how to obtain formula (\ref{energy}).
First one notices
that commutators of supercharges (\ref{salgebra}) form a block
matrix $M_{\alpha \beta}$. The BPS condition means that this matrix has a zero eigenvector. As the block
$\{\tilde Q_{\alpha}, \tilde Q_{\beta}\}$ is non-degenerate one can reformulate the BPS  condition
as the  degeneracy condition on some square matrix of dimension twice smaller than  that of
$M_{\alpha \beta}$. At that point one can apply the
standard technique of finding zero eigenvalues of matrices expressed in terms of Gamma matrices
(for example see \cite{OP} Appendix B).

Note that operators ${\cal P}_{\tilde\imath}$, $p^{\tilde\jmath}$ have a continuum spectrum. Everywhere below we  restrict ourselves to the zero eigenvalue subspace for these operators. Moreover, we will not 
consider any effects of scalar fields on the spectrum.  Below 
${\cal P}_{i}$ denotes the operator  $ {\rm Tr} F_{ij}P^{j}$. 


The phase space $\cal P$ of the theory at hand is not simply connected. It is well known that in such a 
case there is some freedom in  quantization. Namely, in the framework of geometric quantization 
(see \cite{Kirillov} or Appendix A of paper \cite{KS} for
 details) we can assign  to every function $F$ on a phase space an operator $\check F$ defined
 by the formula
\begin{equation} \label{preq}
\check F \phi = F\phi + \omega^{ij}\frac{\partial F}{\partial x^{j}} 
\nabla_{i} \phi 
\end{equation} 
where  $\nabla_{i} \phi = \hbar \partial_{i} + \alpha_{i}$ is a 
covariant derivative with respect to $U(1)$-gauge field having the 
curvature $\omega_{ij}$. Here $\omega_{ij}$ is a matrix of symplectic form $\omega$ and $\omega^{ij}$ 
is the inverse matrix. The operator $\check F$ acts on the space of sections of a line bundle over 
the phase space. The transition from $F$ to $\check F$ is called prequantization. It satisfies 
$$ 
[\check F, \check G ] = i\hbar (\{ F, G \} )^{\vee}
$$
where $\{ F, G \} = \frac{\partial F}{\partial x^{i}}\omega^{ij} 
\frac{\partial G}{\partial x^{j}} $ is the Poisson bracket.
Notice that replacing $\alpha$ with $\alpha + \delta \alpha$ we change $\check F$ in the following 
way
\begin{equation} \label{F}
\check F = \check F_{old} + \delta \alpha (\xi_{F}) = 
\check F_{old} + \delta \alpha_{i}\xi^{i}_{F}
\end{equation}  
where $\xi_{F}$ stands for the Hamiltonian vector field corresponding to the function $F$. 
To define a quantization we should introduce the so called polarization, i.e. exclude half of the 
variables. Then for every function $F$ on the phase space we obtain a quantum operator $\hat F$.

If the gauge field in (\ref{preq}) is replaced by a gauge equivalent field we obtain an equivalent 
quantization procedure. However, in the case when the phase space is not simply connected the gauge 
class of $\nabla_{i}$ in (\ref{preq}) is not specified uniquely. The simplest choice of the 1-form 
$\alpha$ in the system we consider is 
\begin{equation} \label{alpha}
\alpha = {\rm Tr}  P^{I} \delta \nabla_{I}
-i\frac{V}{2g_{YM}^{2}}{\rm Tr} \bar \psi \Gamma^{0}  \delta \psi \, .
\end{equation}
But we can also add to this $\alpha$ any closed 1-form $\delta \alpha$, for example 
\begin{equation} \label{dalpha}
\delta \alpha = \lambda^{i}{\rm Tr} \delta \nabla_{i} + \lambda^{ijk}{\rm Tr}\delta \nabla_{i}
\cdot F_{jk} 
\end{equation}
where $\lambda^{ijk}$ is an antisymmetric 3-tensor and $\lambda^{i}$ is a 1-tensor.
One can check that for $d\le 4$  we obtain all gauge classes by adding (\ref{dalpha}) 
to (\ref{alpha}) (for $d>4$ one 
should include additional terms labeled by antisymmetric tensors of odd rank $\ge 5$).
In the Lagrangian formalism  one can include topological terms into the action functional.
The addition of all topological terms to the Lagrangian corresponds to the consideration 
of all possible ways of (pre)quantization in the Hamiltonian formalism.
The topological terms can be interpreted as Ramond-Ramond backgrounds from the string theory 
point of view (see \cite{HofVer2}, \cite{BrMorZum2} for more details).
The additional terms  in the action can be expressed in terms of the Chern character.
In particular, the consideration of the form $\alpha + \delta \alpha$ given by  
(\ref{alpha}), (\ref{dalpha}) corresponds to adding the following
 terms to the action
\begin{equation}\label{Stop}
S_{top} = \lambda^{i}{\rm Tr} F_{0i} + \lambda^{ijk}{\rm Tr} F_{0i}F_{jk}  \, .
\end{equation}

Let us make one  general remark here. Assume that a first order linear differential operator $L_{A}$
defined by $L_{A}F = \{A,G\}_{PB}$ obeys $e^{2\pi  L_{A}}=1$ (in other words the Hamiltonian vector field 
corresponding to $A$ generates the group $U(1)={\rm \bf R}/{\rm \bf Z}$). Then it is easy to 
check that 
\begin{equation}\label{A}
e^{2\pi  \check A} \check F e^{-2\pi  \check A} = \check F
\end{equation} 
for any function $F$ on the phase space. 
It follows from (\ref{A}) that after quantization we should expect 
$$
e^{2\pi i \hat A}=const. 
$$
We see that the eigenvalues of $\hat A$ are quantized; they have the form $m + \mu$ 
where $m\in {\rm \bf Z}$ and $\mu$ is a fixed constant.

 Now let us  discuss the spectrum of the operators  $\hat {\cal P}_{i}$, $\hat p^{i}$.
Our considerations will be along the line  proposed in \cite{HofVer2}. 
Consider a Hamiltonian vector field $\xi_{{\cal P}_{i}}$ corresponding to the gauge invariant momentum functional ${\cal P}_{i} = {\rm Tr}F_{ij}P^{j}$. The variation of $\nabla_{j}$ under the action 
of $\xi_{{\cal P}_{i}}$  is equal to 
$F_{ij}$. Let us fix a connection $\nabla^{0}_{i}$ on the module $E$. 
Then an arbitrary connection has the form $\nabla_{i} = \nabla_{i}^{0} + X_{i}$ where 
$X_{i}\in End_{T_{\theta}}E$. Hence, we have
\begin{equation} \label{P}
\xi_{{\cal P}_{i}}(\nabla_{j}) = F_{ij} = -[\nabla_{j}, X_{i}] +  [\nabla_{i}^{0},\nabla_{j}] \, . 
\end{equation} 
Here the first term corresponds to infinitesimal gauge transformation. Therefore, up to a 
gauge transformation, ${\cal P}_{i}$ generates a vector field acting as 
$\delta \nabla_{j} =  [\nabla_{i}^{0},\nabla_{j}]$. Now one can check using 
 (\ref{delta}), (\ref{Leib}) and the identity $e^{ \delta_{j}}=1$ 
that the operator $exp( \nabla^{0}_{j})$ is an endomorphism, i.e. 
$$
exp(  \nabla^{0}_{j}) U_{i} exp(-  \nabla^{0}_{j}) = U_{i}
$$
and since it is  unitary it  represents a global gauge transformation.
This means that on the phase space of our theory (after taking a quotient with respect to the 
gauge group) the identity $exp(  L_{{\cal P}_{j}})=1$ is satisfied. Here $L_{{\cal P}_{j}}$
stands for the differential operator corresponding to ${\cal P}_{j}$. 
As we already mentioned this identity leads to a quantization of $\hat {\cal P}_{j}$ eigenvalues. 
Namely, for some fixed constant $\mu_{j}$ the eigenvalues of $\hat {\cal P}_{j}$ have the form 
\begin{equation} \label{qc}
{\cal P}_{j} = 2\pi( \mu_{j} + m_{j}) 
\end{equation}
where $m_{j}$ is an integer. 
The quantization condition (\ref{qc}) is valid for every choice of quantization procedure
(for every $\alpha + \delta \alpha$). However, the constants $\mu_{j}$ depend on 
this choice. Using the freedom that we have in the quantization we can take $\mu_{j}=0$ 
when $\delta \alpha=0$. In the case of non-vanishing $\alpha$ we obtain using (\ref{F}) and (\ref{P})
\begin{equation}
 \label{momentum}
 {\cal P}_{j} = 2\pi m_{j} +  C_{jk}\lambda^{k} + C_{ijkl}\lambda^{jkl} \, .
\end{equation}

 Quantization of the electric field zero mode ${\rm Tr}P^{i}$ comes from periodicity conditions on
the space $Conn' E/G$. 
It was first noted in \cite{HofVer2} one can consider
the operator $U_{j}exp(-  \theta^{jk}\nabla_{k})$ that commutes with all $U_{i}$ and thus is
an endomorphism. Being unitary this operator determines a gauge transformation. As one can easily 
check this gauge transformation can be equivalently written as an action of the 
operator 
\begin{equation} \label{op}
exp((2\pi \delta_{jk} - \theta^{jl}F_{lk})\frac{\delta}{\delta\nabla_{k}})
\end{equation}
on the space $Conn'E$. Therefore, on the space $Conn' E/G$ this operator descends to the  identity. 
The Hamiltonian vector field defined by the exponential of (\ref{op}) corresponds to the functional 
$2\pi p^{k} - \theta^{ki}{\cal P}_{i}$. Hence, as we explained above the quantum operator 
$\hat p^{k} - (2\pi)^{-1}\theta^{ki}\hat {\cal P}_{i}$ has eigenvalues of the form 
$ n^{i} + \nu^{i}$ 
where $n^{i} \in {\rm \bf Z}$ and $\nu^{i}$ is a fixed number. 
  Proceeding as above we obtain the following quantization law for eigenvalues of $Tr\hat P^{i}$
\begin{equation} \label{p}
p^{i} =  n^{i} + \theta^{ij}m_{j} +  \lambda^{i}dimE +
\lambda^{ijk} C_{jk}
\end{equation}
where $n^{i}$ is an integer and $m_{j}$ is the integer specifying the eigenvalue of total momentum 
operator (\ref{momentum}).

\section{Energies of BPS states for $d=2,3,4$}
Now we are ready to write explicit answers for $d=2,3,4$. For
$d=2$ the Chern character can be written as
$$
ch(E) = (p - q\theta) + q\alpha^{1}\alpha^{2}
$$
where $p$ and $q$ are integers such that $p-q\theta > 0$.
 Hence $dimE=p-q\theta$, $C_{ij} = \pi \epsilon_{ij} q$, $C_{2}=0$ and we get the
following answer for the energies of BPS states
\begin{eqnarray} \label{d=2}
&&E= \frac{R_{0}^{2}g_{YM}^{2}}{2VdimE}(n^{i} +\theta^{ik}m_{k}+  \lambda^{i}dimE )g_{ij}\cdot
\nonumber \\
&&\cdot (n^{j} + \theta^{jl}m_{l}+  \lambda^{j}dimE) +  \nonumber \\
&& + \frac{R_{0}^{2}}{2Vg_{YM}^{2}dimE}(\pi q +
 \phi dimE)^{2}  +  \frac{2\pi R_{0}}{dimE}\| { \bf v }\|
\end{eqnarray}
where $ \bf $ is a two-dimensional vector
${\bf v} = (m_{i}p-q\epsilon_{ij}n^{j} )$.

For a three-dimensional  torus the Chern character has the form
$$
ch(E) = p + \frac{1}{2}{\rm tr}\theta q  + \frac{1}{2}q_{ij}\alpha^{i}\alpha^{j}
$$
where $p$ is an integer and $q$ is an antisymmetric matrix with integral entries.
Substituting the expressions $dimE= p + \frac{1}{2}{\rm tr}\theta q$,
$C_{ij} = \pi q_{ij}$, $C_{2}=0$ into the main formula (\ref{energy}) and using (\ref{p}), (\ref{momentum})
we obtain
\begin{eqnarray} \label{d=3}
&&E= \frac{R_{0}^{2}g_{YM}^{2}}{2VdimE}(n^{i} + \theta^{ik}m_{k}+ \lambda^{i}dimE + 
\pi \lambda^{ikl}q_{kl})g_{ij}\cdot \nonumber \\
&& \cdot (n^{j} + \theta^{jr}m_{r} + \lambda^{j}dimE + \pi \lambda^{jrs}q_{rs}) +  \nonumber \\
&& + \frac{V}{4g_{YM}^{2}dimE}(\pi q_{ij} + dimE \phi_{ij})g^{ik}g^{jl}(\pi q_{kl} + dimE \phi_{kl})
+ \nonumber \\
&& + \frac{\pi R_{0}}{dimE}\| {\bf v}\|
\end{eqnarray}
where ${\bf v} = (m_{i}dimE - q_{ij}(n^{j} + \theta^{jk}m_{k}))$.

Finally let us consider the case $d=4$. The Chern character now reads as
$$
ch(E) = p + \frac{1}{2}{\rm tr}\theta q + s\underline{\theta} +
\frac{1}{2}(q + *\theta s)_{ij}\alpha^{i}\alpha^{j} + s\alpha^{1}\alpha^{2}\alpha^{3}\alpha^{4}
$$
where $p$ and $s$ are integers, $q_{ij}$ is an antisymmetric $4\times 4$ matrix with integral
entries, $(*\theta)_{ij}=\frac{1}{2}\epsilon_{ijkl}\theta^{kl}$, and 
$\underline{\theta}=\frac{1}{8}\theta^{ij}\epsilon_{ijkl}\theta^{kl} = \sqrt{det \theta_{ij}}$ 
stands for the Pfaffian of $\theta$. From this expression for $ch(E)$   we obtain
$$
dimE=p + \frac{1}{2}{\rm tr}\theta q + s\underline{\theta} \, , \quad
C_{ij} = \pi (q + *\theta s)_{ij}
$$
$$
C_{[ijkl]} = \pi^{2}\epsilon_{ijkl} s \, , \quad C_{2} = (dimE)^{-1}(ps - \underline{q}) \, .
$$
Substituting the above expressions  into (\ref{energy}) we obtain the following answer for the
BPS spectrum
\begin{eqnarray}\label{d=4}
&&E= \frac{R_{0}^{2}g_{YM}^{2}dimE}{2V}(n^{i} + \lambda^{i}dimE + \pi \lambda^{ikl}(q+*\theta s)_{kl} + \theta^{ik}m_{k})g_{ij}\cdot \nonumber \\
&& \cdot (n^{j} + \lambda^{j}dimE + \pi\lambda^{jrt}(q+*\theta s)_{rt} + \theta^{jr}m_{r}) + \nonumber \\
&& +  \frac{V\pi^{2}}{4g_{YM}^{2}dimE}((q+*\theta s)_{ij} + dimE \frac{\phi_{ij}}{\pi})g^{ik}\cdot 
\nonumber \\ && \cdot g^{jl}((q+*\theta s)_{kl} + dimE \frac{\phi_{kl}}{\pi})
+ \nonumber \\
&& + \frac{\pi R_{0}}{dimE}\sqrt{ \|{\bf v}\|^{2} + (\pi /g^{2}_{YM})^{2}(ps-\underline{q})^{2} } \, .
\end{eqnarray}
Here
$$
{\bf v} = (v_{i}) = (m_{i}dimE - (q + *\theta s)_{ij}(n^{j} + \theta^{jk}m_{k}) +
2\pi (*\lambda_{3})_{i}(ps-\underline{q}))
$$, and $(*\lambda_{3})_{i}=\frac{1}{3!}\epsilon_{ijkl}\lambda^{jkl}$.

\section{Morita equivalence}
As it was first shown in \cite{ASMorita} Yang-Mills theories on noncommutative tori
$T_{\theta}$ and $T_{\hat \theta}$ that are related by (complete) Morita equivalence are physically
equivalent. We refer the reader to papers \cite{ASMorita}, \cite{ASRieffel}, \cite{MZum}
for a rigorous definition and discussion of Morita equivalence and its relation to physical
duality. Here we will only state the basic results. Compactifications of M(atrix) theory
  on two $d$-dimensional
noncommutative  tori
$T_{\theta}$ and $T_{\hat \theta}$ are physically equivalent iff
\begin{equation}\label{Morita}
\hat \theta = (M\theta + N)(R\theta + S)^{-1}
\end{equation}
where $M$, $N$, $R$, $S$ are $d\times d$ blocks of the $2d\times 2d$ matrix
\begin{equation}\label{g}
\left(
\begin{array}{cc}
M&N\\
R&S\\
\end{array}
\right)
\end{equation}
belonging to the group $SO(d,d|{\rm \bf Z})$.
Correspondence between different quantities defined on a module $E$ over $T_{\theta}$ and
on a module $\hat E$ over $T_{\hat \theta}$ specified by Morita equivalence, can be explicitly
described by means of  the matrix (\ref{g}). The metric $g_{ij}$ transforms in the following way
\begin{equation} \label{metric}
\hat g = AgA^{t}
\end{equation}  where $A=S + R\theta$. For the curvatures one has the relation
\begin{equation}\label{curv}
AF^{\nabla}A^{t}  - \pi RA^{t} = F^{\hat \nabla} \, .
\end{equation}
 The antisymmetric tensor $\phi_{ij}$ transformation reads
\begin{equation} \label{phi}
\hat \phi = A\phi A^{t} + \pi RA^{t} \, .
\end{equation}

The dimensions of $E$ and $\hat E$ are connected by the relation
\begin{equation}\label{dim}
dim\hat E = dimE|detA|^{-1/2} \, .
\end{equation}
Using (\ref{alpha}) or (\ref{Stop}) and (\ref{curv}) one can easily find transformation laws for the tensors $\lambda^{i}$,
$\lambda^{ijk}$ under Morita equivalence (\ref{Morita}):
\begin{equation}\label{l3hat}
\hat \lambda^{ijk} =|detA|^{1/2} B^{i}_{l}B^{j}_{m}B^{k}_{n} \lambda^{lmn}
\end{equation}
\begin{equation}\label{l1hat}
\hat \lambda^{i} = |detA|^{1/2}B^{i}_{j}\lambda^{j} + \hat \lambda^{ijk}\sigma_{jk}
\end{equation}
where $\sigma = -\pi RA^{t}$, $B=(A^{t})^{-1}$.

The  eigenvalues $p^{i}$ and ${\cal P}_{i}$ transform as follows
\begin{eqnarray}\label{pp}
&&\hat p^{i} = {(A^{t})^{-1}}^{i}_{j}p^{j} \, , \nonumber \\
&& \hat {\cal P}_{i} = A_{i}^{j}{\cal P}_{j} - \pi R_{ij}p^{j}
\end{eqnarray}

Using the above formulas it is easy to check the invariance of spectrum (\ref{energy}) provided
the Yang-Mills  coupling constant $g_{YM}^{2}$ transforms according to the formula
$\hat g_{YM}^{2} = |det A|^{1/2}g_{YM}^{2}$. One can also check the invariance of
expressions (\ref{d=2}), (\ref{d=3}), (\ref{d=4}) written in terms of topological numbers under
 $SO(d,d|{\rm \bf Z})$ transformations. One just needs to note that, as it is 
shown in \cite{ASMorita}, the element $\mu(E)$
of a Grassmann algebra containing topological numbers 
transforms according to a spinor representation of $SO(d,d|{\rm \bf Z})$ under Morita equivalence transformation (\ref{Morita}).

\section{Modules with constant curvature connections}

In the previous sections we calculated the energies of BPS states in the assumption that these states 
do exist. In the present section we will confirm our calculations by means of explicit semiclassical 
construction. We will consider 1/2 BPS fields (constant curvature connections).
 One can calculate 
the energies of corresponding BPS-states restricting ourselves to a neighborhood of the 
set of constant curvature connections. For $d=2$ 
the  calculations of this kind were performed in  \cite{KS}. The answer agrees  
with the calculations made above. We will show that this result can be extended to the case 
of $d>2$. It is proved in \cite{KS} that under certain conditions on $\theta$ every module 
admitting a constant curvature connection can be transformed into a free module by means of 
$SO(d,d|{\rm \bf Z})$ duality  transformation. This means that it is sufficient to consider 
free modules when calculating BPS energies.

Let $T_{\theta}$ be a noncommutative d-dimensional torus and $E$ a module over it admitting a 
constant curvature connection. 
We assume that $End_{T_{\theta}}E$ is a noncommutative torus $T_{\tilde \theta}$.  Then 
$\tilde \theta$ and $\theta$ are related by means of an element of $SO(d,d|{\rm \bf Z})$ 
specified by a matrix 
\begin{equation} \label{g1}
g=\left(
\begin{array}{cc}
M&N\\
R&S
\end{array}
\right) \, .
\end{equation}
This element of $SO(d,d|{\rm \bf Z})$ transforms our module into a free module of rank 1 
(i.e. the module we consider is a basic module in the terminology of \cite{KS}).
The torus $T_{\tilde \theta}$ has generators $Z_{i}$ satisfying 
$$
Z_{j}Z_{k}=e^{2\pi i \tilde \theta^{jk}}Z_{k}Z_{j} \, . 
$$
Let us consider the set $\cal C$ of all constant curvature connections. The group of 
all gauge transformations acts on this set. This group is disconnected. 
The group of its connected components is denoted by $G^{large}$. One can check that the 
components of the set $\cal C$ are labeled by elements of $G^{large}/G^{mon}$ where 
$G^{mon}$ corresponds to gauge transformations defined by monomials 
$Z_{1}^{k_{1}}Z_{2}^{k_{2}}\cdot \dots \cdot Z_{d}^{k_{d}}$. 
More precisely, it is easy to check that the gauge transformation corresponding to such a monomial 
transforms a constant curvature connection $\nabla^{0}_{i}$ into a connection of the form 
$\nabla_{i}^{0} + c_{i}\cdot {\bf 1}$ that obviously belongs to the same component of $\cal C$. 
One can prove that a gauge transformation belonging to $G^{large}/G^{mon}$ transforms 
$\nabla^{0}_{i}$ into a connection from another component of $\cal C$. 
The proof is based on the reduction to the case of a free module by means of 
$SO(d,d|{\rm \bf Z})$-transformation. 
Now we can describe the part of the phase space corresponding to a small neighborhood of 
the space $\cal C$. We should fix a constant curvature connection $\nabla^{0}$ and consider
small fluctuations of connections of the form $\nabla^{0}_{i} + c_{i}\cdot {\bf 1}$ satisfying the 
Gaussian constraint $[\nabla_{i}, P^{i}]=0$ or more precisely obeying the linearization of this 
identity. (In other words  we consider a small neighborhood of the set 
$\{\nabla_{i}^{0} + c_{i}\cdot {\bf 1} \}$ in the  space $\cal N$ defined above.) 
We should impose also the gauge condition restricting ourselves to fluctuations that are 
orthogonal to the gauge group orbits (see \cite{KS} for details). Then the phase space can be 
obtained from this neighborhood by means of factorization with respect to $G^{mon}$.
(Again the easiest way to verify this statement is to use a $SO(d,d|{\rm \bf Z})$ transformation to 
a free module.)  

Factorization with respect to the group 
$G^{mon}$  leads to the  described above quantization conditions on the eigenvalues of the 
operators  ${\rm Tr} \hat P^{i}$ and $\hat {\cal P}_{j}$. 
Namely,  one first introduces a basis in the Lie algebra $L$ in which 
the condition $[\nabla_{j}, Z_{k}] = 2\pi i\delta_{jk}Z_{k}$ is satisfied. This basis is related 
to the standard basis defined by the relations $[\nabla_{j}, U_{k}]=2\pi i\delta_{jk}U_{k}$ by 
means of the matrix $A=R\theta + S$. In this new basis factorization over the action 
of the group $G^{mon}$ gives the following eigenvalues of the zero mode of electric field 
\begin{equation}\label{e}
\tilde p^{i} =  e^{i} + \tilde  \theta^{ij}\sum_{\bf k} \sum_{m=1}^{d-1}N_{m}({\bf k})k_{j} \, .
\end{equation}
Here $N_{m}({\bf k})$ stands for the energy level of the $m$-th transverse oscillator with 
momentum $\bf k$, $e^{i}$ are integers, $\tilde p^{i}$ denotes  the components calculated with 
respect to the new basis.  Going to the standard basis, i.e. acting by the matrix $A$ on (\ref{e}), 
one obtains   
$$
p^{i} = n^{i} + \theta^{ij}m_{j}
$$
where 
\begin{equation} \label{n^i}
n^{i} = (S^{t})^{i}_{j}e^{j} - (N^{t})^{ij}\sum_{r=1}^{d-1}\sum_{\bf k} k_{j}N_{r}({\bf k}) \, ,
\end{equation}
\begin{equation}\label{m_i}
 m_{i} = (M^{t})_{i}^{j}\sum_{r=1}^{d-1}\sum_{\bf k} k_{j}N_{r}({\bf k}) - R^{t}_{ij}e^{j} 
\end{equation}
where matrices $M$, $N$, $R$, $S$ are the corresponding blocks of the matrix (\ref{g1}) introduced above.
As one can check directly the numbers $m_{i}$ given by formula (\ref{m_i}) are eigenvalues 
of the operator $(2\pi )^{-1}{\cal P}_{i}$ as it is expected from considerations above. 
Furthermore, we obtain that the spectrum in our approximation reads as
\begin{eqnarray} \label{BPSspec}
&& E =  \frac{g_{YM}^{2}R_{0}^{2}}{2VdimE}
(e^{i} + \tilde \theta^{ir}\sum_{\bf k}\sum_{m=1}^{d-1}
N_{m}({\bf k})k_{r})){A_{i}}^{l}
g_{lm}\cdot \nonumber \\
&& \cdot {(A^{t})_{j}}^{m}( e^{j} + 
\tilde \theta^{js}\sum_{m=1}^{d-1}\sum_{\bf k}N_{m}({\bf k})k_{s}) + \nonumber \\
&& + \frac{V dimE}{4g_{YM}^{2}} ( (\pi A^{-1}R)_{ij} + \phi_{ij})g^{ik}g^{jl} 
( (\pi A^{-1}R)_{kl} + \phi_{kl}) + \nonumber \\
&& + (\pi R_{0})\sum_{m=1}^{d-1}\sum_{\bf k} N_{m}({\bf k}) \left( k_{i}((A^{t})^{-1}g^{-1}A^{-1})^{ij}
k_{j} \right)^{1/2}  
\end{eqnarray}
where  $A= R\theta +S$, $dimE = |det A|^{1/2}$. 
One can minimize the energy (\ref{BPSspec}) over the oscillators energy levels  
$N_{r}({\bf k})$ for fixed quantum numbers $n^{i}$ and $m_{j}$. From (\ref{n^i}), (\ref{m_i}) 
one can derive that one needs to minimize over the numbers $N_{r}({\bf k})$ subject to the constraint 
$$
\sum_{r=1}^{d-1}\sum_{\bf k} k_{i}N_{r}({\bf k}) = S_{i}^{j}m_{j} - R_{ij}n^{j} \, . 
$$
When minimizing  it 
suffices to use the fact that the norm of a sum of vectors is always less or equal than the sum of 
 corresponding norms. The minimized BPS-spectrum has the form
\begin{eqnarray} \label{BPSmin}
&&E=\frac{R_{0}( g_{YM})^{2}}{2VdimE}(n^{i} + \theta^{ik}m_{k})g_{ij}(n^{j} + \theta^{jl}m_{l}) 
+ \nonumber \\
&& + \frac{VdimE}{4g_{YM}^{2}}(\pi (A^{-1}R)_{ij} + \phi_{ij})g^{ik}g^{jl}(\pi (A^{-1}R)_{kl} + \phi_{kl})
+ \nonumber \\
&& + R_{0}\pi \sqrt{v_{i}g^{ij}v_{j}}
\end{eqnarray}
where $v_{i}=A^{-1}_{ij}(S_{j}^{k}m_{k} - R_{jk}n^{k})$, $A=R\theta + S$ and $R$ and $S$ 
are blocks of the matrix (\ref{g1}). This formula agrees with formula (\ref{energy}) if one 
takes into account the identity ${\rm Tr}F_{ij}=\pi (A^{-1}R)_{ij}dimE$  
and the fact that the constant $C_{2}$ vanishes for a module admitting a constant curvature 
connection.
We skipped all technical details here as the calculation essentially parallels the one made in 
\cite{KS} for the case $d=2$. We also omitted the possible topological terms in formulas 
(\ref{BPSspec}), (\ref{BPSmin}). One can easily restore them.

\begin{center} {\bf Acknowledgments} \end{center}
We are indebted to K.~Gawedzki, K.~Hofman, B.~Pioline, M.~Rieffel, and B.~Zumino for useful 
discussions.

\end{document}